\documentclass[conference]{IEEEtran}
\IEEEoverridecommandlockouts

\usepackage{cite}
\usepackage{amsmath,amssymb,amsfonts}
\usepackage{algorithmic}
\usepackage{graphicx}
\usepackage{subfig}
\usepackage[font=small]{caption}
\usepackage{textcomp}
\usepackage{xcolor}
\usepackage{soul}

\usepackage{bm}
\usepackage{upgreek}

\usepackage{xurl}

\def\BibTeX{{\rm B\kern-.05em{\sc i\kern-.025em b}\kern-.08em
    T\kern-.1667em\lower.7ex\hbox{E}\kern-.125emX}}

\begin{document}
\setlength{\abovedisplayskip}{6pt}
\setlength{\belowdisplayskip}{6pt}

\thispagestyle{plain}
\pagestyle{plain}

\definecolor{lightgray}{gray}{0.9}
\definecolor{lightblue}{rgb}{0.9,0.9,1}
\definecolor{LightMagenta}{rgb}{1,0.5,1}
\definecolor{red}{rgb}{0.5,1,1}

\newcommand\couldremove[1]{{\color{lightgray} #1}}
\newcommand{\remove}[1]{}
\newcommand{\move}[2]{ {\textcolor{Purple}{ \bf --- MOVE #1: --- }} {\textcolor{Orchid}{#2}} }

\newcommand{\hlc}[2][yellow]{ {\sethlcolor{#1} \hl{#2}} }
\newcommand\note[1]{\hlc[SkyBlue]{-- #1 --}} 

\newcommand\tingjun[1]{\hlc[yellow]{TC: #1}}
\newcommand\yiming[1]{\hlc[pink]{YL: #1}}
\newcommand\michael[1]{\hlc[red]{ML: #1}}
\newcommand\zhihui[1]{\hlc[LightMagenta]{ZG: #1}}

\newcommand{\myparatight}[1]{\vspace{0.5ex}\noindent\textbf{#1~~}}

\newcommand{\greenyes}{\color[HTML]{3C8031}{\textbf{Yes}}}
\newcommand{\redno}{\color[HTML]{ED1B23}{\textbf{No}}}

\newcommand{\specialcell}[2][c]{%
  \begin{tabular}[#1]{@{}c@{}}#2\end{tabular}}

\newcommand{\myCodeShort}[1]{\texttt{\small{#1}}}

\newcommand{\longLeft}{\phi_{l}}
\newcommand{\longRight}{\phi_{r}}
\newcommand{\latTop}{\psi_{t}}
\newcommand{\latBtm}{\psi_{b}}

\newcommand{\resolution}{r}

\newcommand{\area}{\textbf{A}}
\newcommand{\areaSet}{\mathcal{A}}
\newcommand{\areaSizeX}{L_{x}}
\newcommand{\areaSizeY}{L_{y}}
\newcommand{\mapSizeX}{N_{x}}
\newcommand{\mapSizeY}{N_{y}}

\newcommand{\powerTx}{\textit{P}_{\textrm{TX}}}
\newcommand{\powerRx}{\textit{P}_{\textrm{RX}}}
\newcommand{\gainTx}{\textit{G}_{\textrm{TX}}}
\newcommand{\gainRx}{\textit{G}_{\textrm{RX}}}
\newcommand{\lossInsert}{\textit{IL}}

\newcommand{\loss}{L}

\newcommand{\pathGain}{\textit{PG}}

\newcommand{\mapPathGain}{\textbf{P}}

\newcommand{\mapPathGainITM}{\textbf{P}_{\textrm{ITM}}}
\newcommand{\mapPathGainUMa}{\textbf{P}_{\textrm{UMa}}}

\newcommand{\mapPathGainIso}{\textbf{P}_{\textrm{iso}}}
\newcommand{\mapPathGainIsoEst}{\widehat{\textbf{P}}_{\textrm{iso}}}
\newcommand{\mapPathGainIsoSet}{\mathcal{P}_{\textrm{iso}}}

\newcommand{\mapPathGainDir}{\textbf{P}_{\textrm{dir}}}
\newcommand{\mapPathGainDirSet}{\mathcal{P}_{\textrm{dir}}}

\newcommand{\mapBuilding}{\textbf{B}}
\newcommand{\mapBuildingSet}{\mathcal{B}}

\newcommand{\mapSigStrength}{\textbf{S}}
\newcommand{\mapSigStrengthEst}{\widehat{\textbf{S}}}
\newcommand{\mapSigStrengthSparse}{\textbf{S}^{\downarrow}}

\newcommand{\mapRSRP}{\textbf{S}_{r}}
\newcommand{\mapRSRPEst}{\widehat{\textbf{S}}_{r}}
\newcommand{\mapRSRPSparse}{\textbf{S}_{r}^{\downarrow}}

\newcommand{\mapSparse}{\textbf{S}}

\newcommand{\mapStrengthIso}{\textbf{R}_{\textrm{iso}}}
\newcommand{\mapStrengthIsoEst}{\widehat{\textbf{R}}_{\textrm{iso}}}
\newcommand{\mapStrengthDir}{\textbf{R}_{\textrm{dir}}}
\newcommand{\mapStrengthDirEst}{\widehat{\textbf{R}}_{\textrm{dir}}}

\newcommand{\paramIso}{\bm{\uptheta}_{\textrm{iso}}}
\newcommand{\paramDir}{\bm{\uptheta}_{\textrm{dir}}}

\newcommand\name{{\sc Geo2SigMap}}
\newcommand\namebf{{\sc\textbf{Geo2SigMap}}}

\title{{\name}: High-Fidelity RF Signal Mapping Using Geographic Databases}

\author{
\IEEEauthorblockN{Yiming Li, Zeyu Li, Zhihui Gao, Tingjun Chen}
\IEEEauthorblockA{Department of Electrical and Computer Engineering, Duke University \\
Email: \{yiming.li416, zeyu.li030, zhihui.gao, tingjun.chen\}@duke.edu}
}

\maketitle

\begin{abstract}
Radio frequency (RF) signal mapping, which is the process of analyzing and predicting the RF signal strength and distribution across specific areas, is crucial for cellular network planning and deployment. Traditional approaches to RF signal mapping rely on statistical models constructed based on measurement data, which offer low complexity but often lack accuracy, or ray tracing tools, which provide enhanced precision for the target area but suffer from increased computational complexity. Recently, machine learning (ML) has emerged as a data-driven method for modeling RF signal propagation, which leverages models trained on synthetic datasets to perform RF signal mapping in ``unseen'' areas. However, such methods often require the use of advanced proprietary software for creating synthetic datasets (e.g., ray tracing), or rely on measurements collected from the unseen areas to effectively train the models. 

In this paper, we present {\namebf}, an ML-based framework for efficient and high-fidelity RF signal mapping using geographic databases. First, we develop an automated framework that seamlessly integrates three open-source tools: OpenStreetMap (geographic databases), Blender (computer graphics), and Sionna (ray tracing), enabling the efficient generation of large-scale 3D building maps and ray tracing models. Second, we propose a cascaded U-Net model, which is pre-trained on synthetic datasets and employed to generate detailed RF signal maps, leveraging environmental information and sparse measurement data. Finally, we evaluate the performance of {\namebf} via a real-world measurement campaign, where three types of user equipment (UE) collect over 45,000 data points related to cellular information from six LTE cells operating in the citizens broadband radio service (CBRS) band. Our results show that {\namebf} achieves an average root-mean-square-error (RMSE) of {6.04}\thinspace{dB} for predicting the reference signal received power (RSRP) at the UE, representing an average RMSE improvement of {3.59}\thinspace{dB} compared to existing methods.
\end{abstract}

\section{Introduction}
\label{sec: introduction}

Radio frequency (RF) signal mapping, which is the
process of analyzing and predicting the RF signal strength and distribution across specific areas, is crucial for cellular network planning and deployment~\cite{3gpp_5g_nr_bs_radio_tx_rx, sufyan20235g}. Efficient and accurate RF signal mapping allows network service providers to optimize the placement of base stations (BSs) and antennas to provide the desired coverage and guaranteed quality of service to end users. It can also facilitate the operation of multiple radio access technologies and the co-existence of active and passive users sharing the same or adjacent frequency bands~\cite{gao2019performance, coexist_cbrs_c, souryal2019effect}.

Traditional methods for RF signal mapping utilize analytical or empirical path loss (or path gain) models, such as the Friis free space model~\cite{Friis}, 3GPP urban macro (UMa) model~\cite{3GPPTR38.901}, and Ericsson channel model~\cite{Ericsson}. These models describe the path gain (PG) as a function of various system parameters (e.g., carrier frequency, link distance, and antenna height) with different coefficients that are specified depending on the scenarios, such as urban or rural settings. One limitation of these models is their lack of consideration for the geographic information, such as the 3D building and terrain maps, which hinders the accuracy of RF signal mapping across diverse areas.
On the other hand, ray tracing tools~\cite{WinProp, Siradel, sionnaRT, opal}, also known as ray tracers, offer a more sophisticated approach by simulating the transmission of RF signals through the emission of millions of rays within a given environment and analyzing their reflection and diffraction effects. As a result, ray tracing can provide more accurate RF signal mapping but at the cost of increased complexity and required computational resources. In addition, many advanced ray tracing tools are proprietary and require paid licenses, and thus are not readily available or open-source to the broader research community.

In recent years, machine learning (ML) has emerged as a data-driven approach for modeling RF signal propagation, which leverages models trained on synthetic datasets to facilitate RF signal mapping. These synthetic datasets often contain detailed environmental information, such as 3D building and terrain maps, enabling more accurate radio propagation modeling through ray tracing and ML, especially for areas not previously observed. However, these ML-based methods often require advanced proprietary software for creating synthetic datasets (e.g., for ray tracing~\cite{HuaweiKDD, DeepRay2022}), or rely on collecting comprehensive real-world measurements from the explored areas to effectively train the models (e.g.,~\cite{thrane2019DTU, alimpertis2019city, UAV-UNet-cm}). Although there exists a number of open-source tools for ray tracing and creating 3D building maps, the lack of seamless integration across these tools prevents their use by the research community. Furthermore, the ML models designed for RF signal mapping often lack scalability to incorporate additional real-world information or to adapt to different environments.

In this paper, we present {\name}, \emph{a novel framework designed for high-fidelity and efficient RF signal mapping using geographic databases and ML.}
First, we develop an automated framework that seamlessly integrates three open-source software tools: \emph{OpenStreetMap (OSM)}~\cite{OpenStreetMap}, which is a real-world geographic database, \emph{Blender}~\cite{Blender}, which is a 3D computer graphics tool, and \emph{Sionna}~\cite{sionnaRT}, which is a next-generation Physical layer research tool that includes a differentiable ray tracer. This first-of-its-kind integration enables the efficient generation of large-scale 3D building maps and ray tracing models.
Second, we propose an ML model based on a cascaded U-Net architecture, which employs two cascaded U-Nets to learn the RF signal mapping, represented by the signal strength (SS) map for a given geographical area. Specifically, the first U-Net generates a PG map that embeds the environmental information, and the second U-Net further refines this process and generates the fine-grained SS map by incorporating directivity and link budget information, as well as an additional input of a sparsely sampled SS map sampled across the same area.
This cascaded U-Net model is trained using only synthetic datasets, therefore no real-world measurements are required during the training phase. When the pre-trained model is employed to predict the detailed SS map for a specific area, we incorporate a few field measurements that serve as the sparse SS map input to the second U-Net. Such a design effectively streamlines the model's applicability across different areas and eliminates the need for retraining the the entire model for different geographical settings.

Finally, we evaluate the performance of {\name} via a real-world measurement campaign, where three types of user equipment (UE) collect cellular information from six LTE cells operating in the citizens broadband radio service (CBRS) band ({3.55--3.7}\thinspace{GHz}), deployed on the Duke University West Campus. Using customized Android apps and Python scripts, we collect over 45,000 measurements, each including various key cellular metrics such as the physical cell ID (PCI), reference signal received power (RSRP), and reference signal received quality (RSRQ).
Our results show that {\name} achieves an average root-mean-square-error (RMSE) of {6.04}\thinspace{dB} for predicting the RSRP at the UE across the six LTE cells, representing an average improvement of {3.59}\thinspace{dB} compared to existing RF signal mapping methods that rely on statistical channel models, ray tracing, and ML approaches.

To summarize, the main contributions of this paper include:
\begin{itemize}
\item 
We develop an automated framework that seamlessly integrates three open-source tools that specialize in different domains, including geographic databases (OSM), computer graphics (Blender), and ray tracing (Sionna). This integration enables the efficient generation of large-scale 3D building maps and ray tracing models;
\item 
We propose a cascaded U-Net architecture tailored for high-fidelity RF signal mapping using synthetic building maps and ray tracing datasets. This novel ML model employs a two-stage process that leverages both building maps and link budget information to accurately learn and predict fine-grained RF signal mapping;
\item 
We comprehensively evaluate the performance of {\name} through an extensive measurement campaign, where user-side information are collected by different UE types across six LTE cells operating in the CBRS band. We show that the cascaded U-Net model, which is pre-trained on synthetic datasets, achieves significantly improved signal strength prediction accuracy compared to various baseline methods, in real-world scenarios.
\end{itemize}
Code and datasets for {\name} are open-source~\cite{GitHub_repo}.
\section{Related Work}
\label{sec:related}

\noindent\textbf{RF propagation modeling and SS measurements.}
RF propagation modeling and signal strength (SS) measurements are essential for commercial cellular networks to ensure desired coverage and guaranteed quality of service to the end users in both the {sub-7}\thinspace{GHz} and millimeter-wave band~\cite{narayanan2020first,narayanan2022comparative,raychaudhuri2020challenge,du2020directional}.
Recent works have reported extensive LTE/5G measurements and analysis in the wild~\cite{narayanan2021variegated}, including in urban areas and considering the CBRS band~\cite{simic2015can, chicagoMeasurement}, with a focus on the co-existence between different services and activities~\cite{gao2019performance, coexist_cbrs_c, souryal2019effect}.

\vspace{0.5ex}
\noindent\textbf{Ray tracing tools.}
A diverse range of commercial \emph{licensed} ray tracing software is available, with examples such as WinProp~\cite{WinProp}, iBwave Design~\cite{ibwave-design},  Wireless Insite~\cite{wireless-insite}, Volcano~\cite{Siradel}, and MATLAB~\cite{MatlabRayTracing}, each offering specialized features for advanced signal propagation simulation. 
On the other hand, \emph{open-source} alternatives such as Sionna~\cite{sionnaRT} and Opal~\cite{opal} present a more accessible option, but often with a more limited range of features and potentially lower accuracy.

\vspace{0.5ex}
\noindent\textbf{RF signal mapping and SS prediction.}
There are two main categories of RF signal mapping in the context of SS prediction at the \emph{point-level}~\cite{RandForest-cm, thrane2019DTU, he2018multi, Juang-VAE-GAN, xia2020mmWaveNN_VAE, nguyen2022CBRS} and the \emph{map-level}~\cite{niu2018recnet, xia2019cellular, Levie-normalised-mse, DeepRay2022, UAV-UNet-cm, li2020supreme, wang2021csmc, fida2017zipweave, niu2020TransferLearning}.
Point-level SS prediction aims to predicts the SS at a given location at a time. Such predictions can be achieved using analytial/statistical channels models, ray tracing tools, or by random forest~\cite{RandForest-cm} and ML models that take the input of satellite maps~\cite{thrane2019DTU} and public urban data~\cite{he2018multi}. More advanced ML approaches based on variational autoencoder (VAE)~\cite{Juang-VAE-GAN, xia2020mmWaveNN_VAE} and transfer learning targeting at the CBRS band~\cite{nguyen2022CBRS} have also been studied.
Map-level SS prediction, closer to our work, aims to predict the SS map of an entire area. Recent works for map-level SS prediction have employed ML models based on convolutional neural networks (CNNs)~\cite{niu2018recnet, xia2019cellular} and U-Net~\cite{Levie-normalised-mse, DeepRay2022, UAV-UNet-cm}. The input to the ML model includes various features such as the building/satellite maps~\cite{li2020supreme, wang2021csmc},
population and road map~\cite{wang2021csmc}, and sparse SS measurements~\cite{fida2017zipweave, niu2018recnet}.
SS maps can also be estimated using compressive sensing techniques leveraging spatial and temporal continuity. For example, Bayesian compressive sensing has been used for estimating the indoor SS map in Wi-Fi networks~\cite{yang2016updating}.

Most relevant to this work is PLNet~\cite{HuaweiKDD}, which is a state-of-the-art ML-based method that leverages detailed environmental data and cell specifications to accurately predict SS maps. In particular, PLNet uses a single U-Net architecture, whose input features include building, terrain, and clutter maps as well as antenna height, orientation, and beam pattern information. The U-Net model is trained using synthetic SS maps generated by the \emph{licensed} Siradel SAS software~\cite{Siradel} for accurate signal propagation modeling.

\emph{To the best of our knowledge, this is the first work that:
(i) develops an automated framework integrating open-source geographic databases, computer graphics, and ray tracing tools, and
(ii) integrates a novel cascaded U-Net architecture that achieves significantly improved SS map prediction accuracy compared to various baseline methods.}

\section{System Design}
\label{sec:system-design}

In this section, we present the design of {\name}, which is composed of three key modules. The first module is responsible for generating building maps and 3D meshes generation using OSM and Blender for accurate spatial representation of the physical environment. The second module is dedicated to generating the path gain (PG) maps generation using Sionna, an open-source ray tracing tool. The final module integrates an ML engine based on the U-Net architecture tailored for efficient and precise RF signal mapping.

\begin{figure}[!t]
    \centering
    \includegraphics[width=0.99\columnwidth]{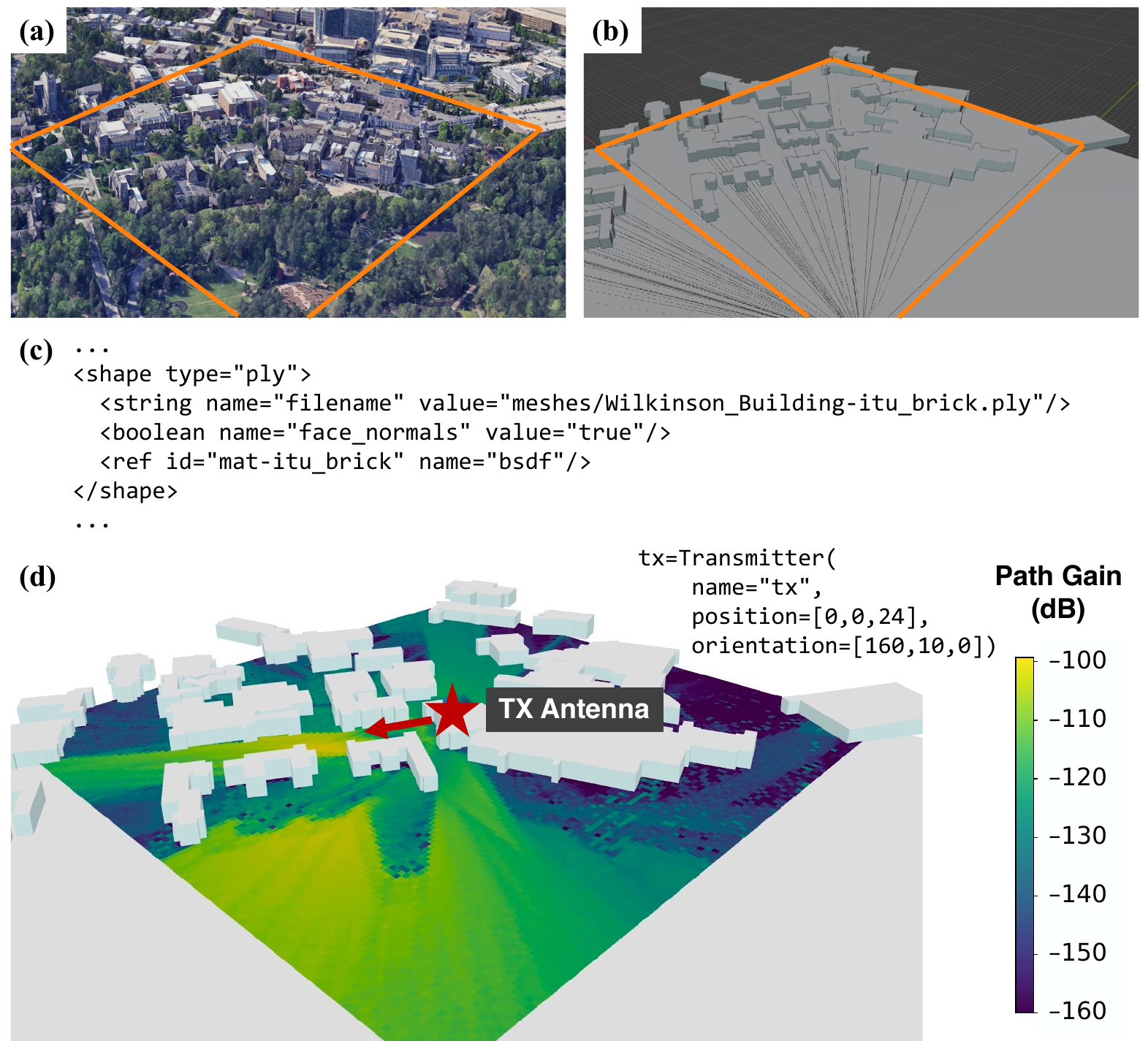}
    \vspace{-3mm}
    \caption{(a) An example {512}\thinspace{m}$\times${512}\thinspace{m} area on the Duke West campus,
    (b) the corresponding building map generated by OSM and rendered by Blender,
    (c) an example building object in the 3D mesh used as input to Sionna, and
    (d) the simulated path gain (PG) map with an antenna placed at the center of the map with a height of {24}\thinspace{m}, {160}$^{\circ}$ azimuth orientation, and {10}$^{\circ}$ downtilt.}
    \label{fig:building-pathgain-map}
\end{figure}

\subsection{Building Maps and 3D Meshes}
\label{ssec:system-design-building-maps}

For a given geographical area $\area$ with dimension $\areaSizeX (\textrm{m}) \times \areaSizeY (\textrm{m})$, we first generate a building map and a 3D mesh that will be used for the ML model and ray tracing simulation.
In particular, the building map is represented by a 1-channel image, $\mapBuilding \in \mathbb{R}^{\mapSizeX \times \mapSizeY}$, with a resolution of $\areaSizeX/\mapSizeX = \areaSizeY/\mapSizeY = \resolution (\textrm{m})$. As a result, the value of each pixel in a building map represents the height of a physical area with dimension $\resolution (\textrm{m}) \times \resolution (\textrm{m})$. We denote the generated synthetic building map dataset by $\mapBuildingSet$.
Fig.~\ref{fig:building-pathgain-map} shows the process to generate the building maps using OSM~\cite{OpenStreetMap}, which is an open geographic database, and Blender~\cite{Blender}, which is an open-source 3D computer graphics software tool.
First, we use OSM to extract the geographic information of a physical area, including the class of the objects (e.g., building, forest) with their corresponding locations and shapes. The extracted information is stored using key-value pairs in the OSM XML format (\myCodeShort{.osm})~\cite{OSM_XML}, where the key describes the context of the location (e.g., buildings or highways) and the value contains the detailed information about the key (e.g., shape and height of a building).
We convert the GPS coordinates (EPSG:4326) used by OSM to Cartesian coordinates that describe the building locations within each area.
Then, based on the building layer data in the exported OSM XML file, we use Blender to model the buildings into 3D objects, which are saved as a 3D mesh in the Polygon File Format (\myCodeShort{.ply}).
We also use the Mitsuba XML format (\myCodeShort{.xml}) to record the material properties of each 3D mesh, such as bricks and glasses. These properties include the objects' relative permittivity and conductivity, from which the reflection/diffraction coefficients are derived and used in the ray tracing process, described next.

\subsection{Ray Tracing-based Path Gain (PG) Maps}
\label{ssec:system-design-ray-tracing}

\begin{table}[!t]
\centering
\caption{Ray tracing parameters employed by Sionna.}
\label{tab:SionnaSpec}
\begin{tabular}{|l|l|}
\hline
\textbf{Parameter} & \textbf{Setting/Value} \\
\hline
Carrier frequency, $f$ & {3.66}\thinspace{GHz} \\
\hline
Area dimension & {512}\thinspace{m}$\times${512}\thinspace{m} \\
\hline
Spatial resolution, $\resolution$ & {4}\thinspace{m} \\
\hline
Reflection & Enabled \\
\hline
Diffraction & Enabled \\
\hline
Maximum \# of reflections/diffractions & 8 bounces \\
\hline
Total \# of rays & 7,000,000 \\
\hline
BS antenna height (area-specific) & $\max(\mapBuilding) + 5 (\textrm{m})$ \\
\hline
BS antenna type & Isotropic, Directional \\
\hline
BS antenna polarization & Dual-polarized (VH) \\
\hline
UE antenan height & {2}\thinspace{m} \\
\hline
UE antenna type & Isotropic \\
\hline
UE antenna polarization & Dual-polarized (VH) \\
\hline
\end{tabular}
\end{table}

\begin{figure*}[!t]
    \centering
    \includegraphics[width=0.99\textwidth]{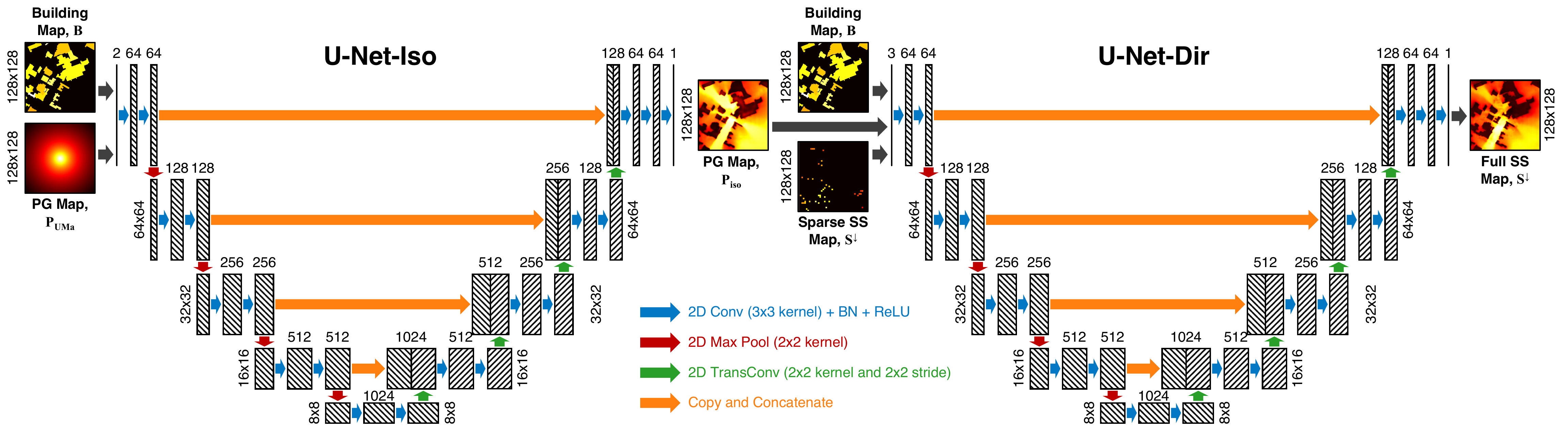}
    \caption{{\name} achieves efficient and precise RF signal mapping via a proposed cascaded U-Net architecture, which is composed of U-Net-Iso and U-Net-Dir for generating coarse path gain (PG) maps and fine-grained signal strengh (SS) maps, respectively.}
    \label{fig:model-cascaded-unet}
\end{figure*}

For each area $\area$, we employ Sionna~\cite{sionnaRT} to generate PG maps for each area based on the 3D mesh. Similar to the building map, $\mapBuilding$, the PG map is represented by a 1-channel image, $\mapPathGain \in \mathbb{R}^{\mapSizeX \times \mapSizeY}$, whose pixel values correspond to the PG values in the dB scale.
For each area, we assume that a BS is located at the center of the area at a height of $\max(\mapBuilding) + 5 (\textrm{m})$, e.g., the BS antenna is {5}\thinspace{m} above the highest point of the building map.
To generate the synthetic PG datasets using realistic building maps, we consider two types of antennas employed by the BS:
(\emph{i}) an \emph{isotropic antenna} with {0}\thinspace{dBi} gain in all directions, and
(\emph{ii}) a \emph{directional antenna} with a boresight gain of {6.3}\thinspace{dBi} and a horizontal/vertical half-power beamwidth (HPBW) of 65$^{\circ}$/8$^{\circ}$, which are typical parameters for directional cellular antennas (e.g., Airspan AirSpeed 1030 in the CBRS band~\cite{AirSpeed}).
For the directional antenna, we consider four orientations randomly selected in the azimuth plane. 
Since our focus is in the CBRS band, we use Sionna to generate the PG maps, $\mapPathGainIso$ and $\mapPathGainDir$, with a carrier frequency at {3.66}\thinspace{GHz} and other parameters summarized in Table~\ref{tab:SionnaSpec}.
In particular, the pixels in each PG map corresponding to the buildings in the area are excluded since we focus on RF signal mapping in outdoor areas. Overall, five PG maps (one $\mapPathGainIso$ and four $\mapPathGainDir$) are generated with respect to each building map $\mapBuilding$ that corresponds to an area $\area$. We denote the generated synthetic PG map datasets by $\mapPathGainIsoSet$ and $\mapPathGainDirSet$, respectively. In Section~\ref{sec:implementation}, we describe the details of the generated PG datasets used for model training.

\subsection{Cascaded U-Net for RF Signal Mapping}
\label{ssec:system-design-cascaded-unet}

At the core of {\name} is an ML engine for high-fidelity RF signal mapping from geographic databases.
In this work, we focus on the reconstruction of signal strength (SS) maps, but the proposed framework can also be extended to predicting other metrics in a given area, such as signal coverage and signal-to-interference-plus-noise ratio (SINR).
The key insight toward the design of the ML model is a two-stage process that turns a building map into a SS map using two cascaded U-Nets, as shown in Fig.~\ref{fig:model-cascaded-unet}.
In particular, the first U-Net (U-Net-Iso) takes inputs of a building map, $\mapBuilding$, and its PG map based on the UMa model, $\mapPathGainUMa$, and generates the \emph{coarse} PG map, $\mapPathGainIso$, assuming an isotropic antenna is used at the BS.
The second \emph{fine-grained} U-Net (U-Net-Dir) takes inputs of the building map, $\mapBuilding$, the coarse PG map, $\mapPathGainIso$, and a sparse SS map, $\mapSigStrengthSparse$, and generates the full SS map for the entire area, $\mapSigStrength$.
In essence, U-Net-Iso learns $\mapPathGainIso$ by adding the knowledge of the building map information to the PG map based on the UMa model, while U-Net-Dir further learns $\mapSigStrength$ by incorporating directivity and link budget information.

The first U-Net, named \emph{U-Net-Iso}, takes a 2-channel image $[\mapBuilding; \mapPathGainUMa] \in \mathbb{R}^{\mapSizeX \times \mapSizeY \times 2}$ as the input and generates the isotropic PG map $\mapPathGainIso$, where $\mapPathGainUMa$ denotes a lightweight PG map based on the 3GPP UMa channel model~\cite{3GPPTR38.901}.
This 2-channel input allows for U-Net-Iso to learn the residual between $\mapPathGainUMa$ and $\mapPathGainIso$ when taking into account the specific building map for the area, therefore facilitating the training process.
We consider a U-Net model consisting of 9 convolutional blocks and 4 downsample/upsample layers. Each convolutional block includes two 2D convolutional layers with a kernel size of {3}$\times${3} and a padding size of {1}$\times${1}, which ensures that the dimensions of input and output remain consistent, and is followed by a 2D batch normalization (BN) layer with ReLU serving as the activation function. In the contracting path (left side), downsampling is achieved by a {2}$\times${2} max pooling operation, whereas in the expansive path (right side), upsampling is achieved by a 2D transposed convolutional layer with {2}$\times${2} stride and {2}$\times${2} kernel size. The number of input/output channels of the convolutional layers are doubled (64$\to$128$\to$256$\to$512$\to$1,024) as the image size decreases by a half (128$\to$64$\to$32$\to$16$\to$8).
Moreover, the output of each convolutional block in the contracting path is copied and concatenated to the input to corresponding convolutional block in the expansive path. This copy and concatenation operation mitigates the potential vanishing/exploding gradients issue. We denote the set of (trainable) parameters in U-Net-Iso across the 2D convolutional layers and BN layers as $\paramIso$.

The second U-Net, named \emph{U-Net-Dir}, employs an identical architecture as U-Net-Iso but with different input and output, as shown in Fig.~\ref{fig:model-cascaded-unet}. In particular, U-Net-Dir takes a 3-channel image, $[\mapBuilding; \mapPathGainIso; \mapSigStrengthSparse] \in \mathbb{R}^{\mapSizeX \times \mapSizeY \times 3}$ as the input, whose 3 channels correspond to the building map, the PG map generated by U-Net-Iso, and a sparse SS map, $\mapSigStrengthSparse$, that includes a small number of SS values sparsely sampled across the area.

The output of U-Net-Dir is the full SS map of the entire area, denoted by $\mapSigStrength \in \mathbb{R}^{\mapSizeX \times \mapSizeY}$. We apply the following link budget equation (in the dB scale) between a transmitter (TX) and receiver (RX) to generate a diverse range of synthetic SS maps based on the directional PG maps, $\mapPathGainDir$,
\begin{align}
    \mapSigStrength\thinspace\textrm{[dBm]} =
    & \powerTx\thinspace\textrm{[dBm]} + \gainTx\thinspace\textrm{[dB]} + \mapPathGainDir\thinspace\textrm{[dB]} \nonumber \\
    & + \gainRx\thinspace\textrm{[dB]} - \lossInsert\thinspace\textrm{[dB]},\ \forall \area,
    \label{eq:signal-strength-link-budget}
\end{align}
where $\powerTx$, $\gainTx$, $\gainRx$, and $\lossInsert$ denote the TX power and gain at the BS, RX gain at the UE, and potential insertion loss of the link, respectively. For each area of the synthetic dataset, the values of $\powerTx$, $\gainTx$, $\gainRx$, and $\lossInsert$ are independently drawn from a specified random distribution (described in Section~\ref{ssec:implementation-dataset-training}), which are then used to generate the SS map for the same area based on {\eqref{eq:signal-strength-link-budget}}. This approach simulates a wide range of BS and UE specifications, alongside other influential factors not accounted for by $\mapPathGainIso$, such as shadowing, obstructions, and UE orientation. In essence, U-Net-Dir is designed to predict the full SS map of a given area based on a few SS values sampled across the area, together with the building map and the PG map produced by U-Net-Iso.
We denote the set of (trainable) parameters in U-Net-Iso across the 2D convolutional layers and BN layers as $\paramDir$.

The cascaded U-Net is trained in a 2-stage procedure using the synthetic datasets $\mapBuildingSet$, $\mapPathGainIsoSet$, and $\mapPathGainDirSet$.
In the first stage, the parameters of U-Net-Iso, $\paramIso$, are trained using the synthetic isotropic PG dataset, $\mapPathGainIsoSet$, with a loss function given by the mean squared error (MSE) between the predicted PG map, $\mapPathGainIsoEst$, and the synthetic PG map, $\mapPathGainIso$, i.e.,
\begin{align}
    \mathcal{L}(\mapBuilding, \paramIso) = \frac{1}{\mapSizeX \mapSizeY} \cdot \big|\big| \mapPathGainIso - \mapPathGainIsoEst(\mapBuilding, \paramIso) \big|\big|_{\textrm{F}}^{2},
\end{align}
where $||\textbf{X}||_{\textrm{F}}$ denotes the Frobenius norm of a matrix given by $||\textbf{X}||_{\textrm{F}} = \sqrt{\sum_{i,j} |X_{i,j}|^{2}}$.
In the second stage, $\paramIso$ is froze and the parameters of U-Net-Dir, $\paramDir$, are trained using the synthetic directional PG dataset, $\mapPathGainDirSet$.
Similarly, we use a loss function given by the MSE between the predicted SS map, $\mapSigStrengthEst$, and ground truth SS map, $\mapSigStrength$, i.e.,
\begin{align}
    \mathcal{L}(\mapBuilding, \mapSigStrengthSparse, \paramDir) = \frac{1}{\mapSizeX \mapSizeY} \cdot \big|\big| \mapSigStrength - \mapSigStrengthEst(\mapBuilding, \mapSigStrengthSparse, \paramDir) \big|\big|_{F}^2.
\end{align}
In Section~\ref{ssec:implementation-dataset-training}, we provide details on the dataset generation and process to train the model parameters $(\paramIso, \paramDir)$.

\section{Implementation and Measurements}
\label{sec:implementation}

\subsection{Dataset Generation and Cascaded U-Net Model Training}
\label{ssec:implementation-dataset-training}

\begin{figure}[!t]
    \centering
    \includegraphics[width=0.98\columnwidth]{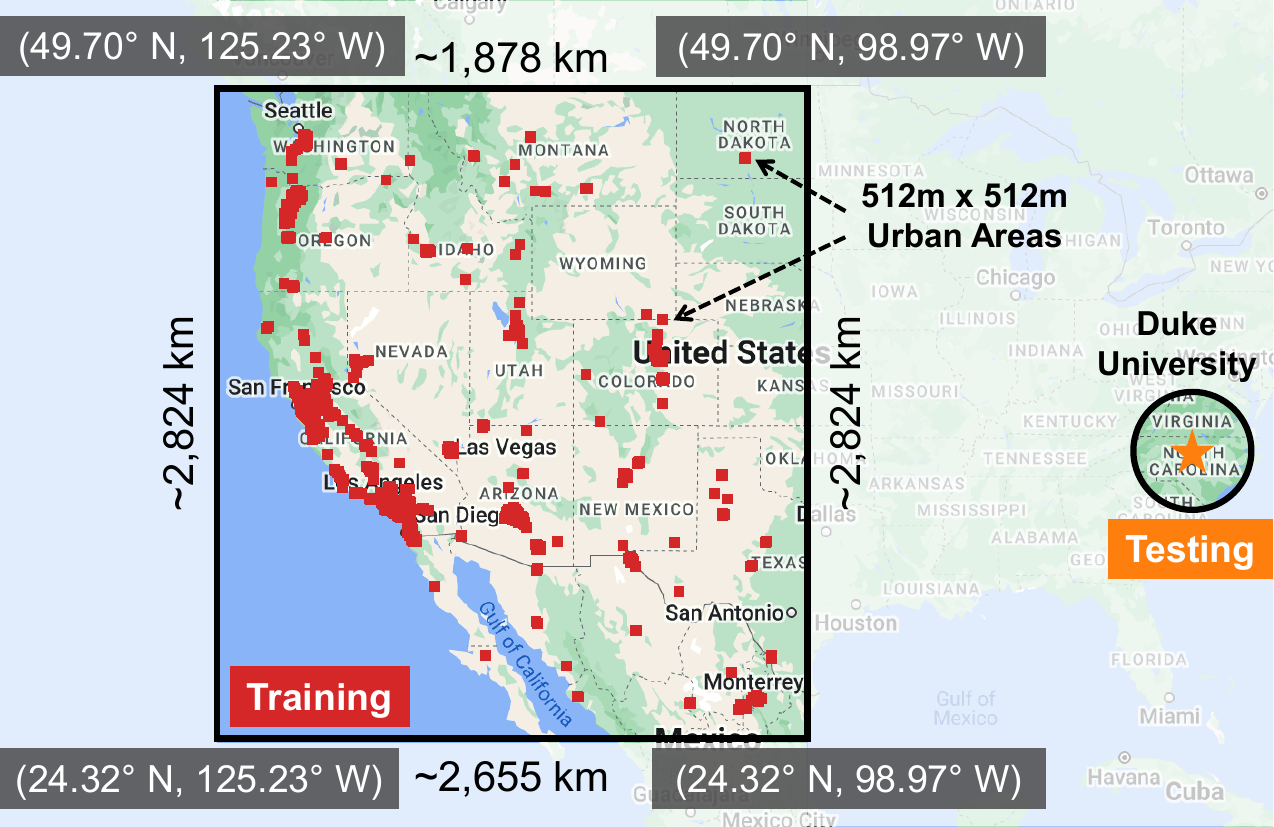}
    \vspace{-1mm}
    \caption{A 6.41 million {km}\textsuperscript{2} area in North America (\emph{left}), from which a total number of 27,176 {512}\thinspace{m}$\times${512}\thinspace{m} areas with a building-to-land ratio of at least 20\% are selected to generate the building map and PG map datasets used to train the cascaded U-Net model in {\name}. The trained model is evaluated using measurements conducted on the Duke University campus (\emph{right}).}
    \label{fig:area-north-america}
\end{figure}

We generate large-scale synthetic building maps and ray tracing datasets used to train the ML model in {\name}. We consider a 6.41 million {km}\textsuperscript{2} area in North America, as shown in Fig.~\ref{fig:area-north-america}, which is divided in to 24.46 million non-overlapping areas with the same dimension of {512}\thinspace{m}$\times${512}\thinspace{m} ($\areaSizeX = \areaSizeY = 512\thinspace\textrm{m}$). Note that the number of areas is larger at lower latitudes. Since we focus on RF signal mapping in urban and suburban areas, we select the areas with a building-to-land ratio of at least 0.2, i.e., at least 20\% of the area is covered by building footprints. As a result, a total number of 27,176 areas are selected, covering a total landscape of {7,124}\thinspace{km\textsuperscript{2}}.

The dataset generation is implemented on Sionna~0.15.1 and Blender~3.3.1, and the code is open-source at~\cite{GitHub_repo}.
For each selected area, we generate one building map ($\mapBuilding$) together with its corresponding 3D mesh, one PG map based on the UMa channel model ($\mapPathGainUMa$), and five PG maps based on Sionna (one $\mapPathGainIso$ and four $\mapPathGainDir$), following the process described in Section~\ref{sec:system-design}.
Each map is represented by a 1-channel 128$\times$128 image ($\mapSizeX = \mapSizeY = 128$) with a spatial resolution of {4}\thinspace{m}, and the parameters used by Sionna are summarized in Table~\ref{tab:SionnaSpec}. Overall, the synthetic dataset consists of 27,176 building maps generated by OSM and 135,880 PG maps generated by Sionna.
To obtain the synthetic SS maps, $\mapSigStrength$ used by U-Net-Dir, we consider the following uniform distributions in the dB scale to generate the parameters to be used by the link budget equation {\eqref{eq:signal-strength-link-budget}} with the PG maps, $\mapPathGainDir$: 
$\powerTx \sim$ {Unif}($+$10, $+$35)\thinspace{dBm}, $\gainTx, \gainRx \sim$ {Unif}(10, 20)\thinspace{dB}, and $\lossInsert \sim$ {Unif}($-$10, $+$10)\thinspace{dB}.

The generated synthetic dataset is split into training and validation sets with a split ratio of 0.8:0.2, and used to train the cascaded U-Net model following the procedure described in Section~\ref{ssec:system-design-cascaded-unet}, with Adam optimizer of learning rate 1e-3 and batch size of 64.
The sparse SS maps are generated as follows to be used during the training phase. For each area in an epoch, we randomly select $N_{\textrm{sparse}}$ points from the outdoor area of the full SS map, $\mapSigStrength$, to form the sparse SS map $\mapSigStrengthSparse$, where $N_{\textrm{sparse}}$ is drawn from the distribution {Unif}[1, 200].

The model is trained using an NVIDIA A100 GPU over 200 epochs, during which the model parameters $(\paramIso, \paramDir)$ with the lowest loss on the validation set are selected. We also apply data augmentation in the training phase by rotating (0$^{\circ}$/90$^{\circ}$/180$^{\circ}$/270$^{\circ}$) and mirroring both the input and output maps (images), which enlarges the dataset by 8$\times$. Overall, the cascaded U-Net model trained using only synthetic datasets consists of 31.04 million parameters.

\begin{figure*}[!t]
    \centering
    \subfloat[]{
    \includegraphics[height=1.88in]{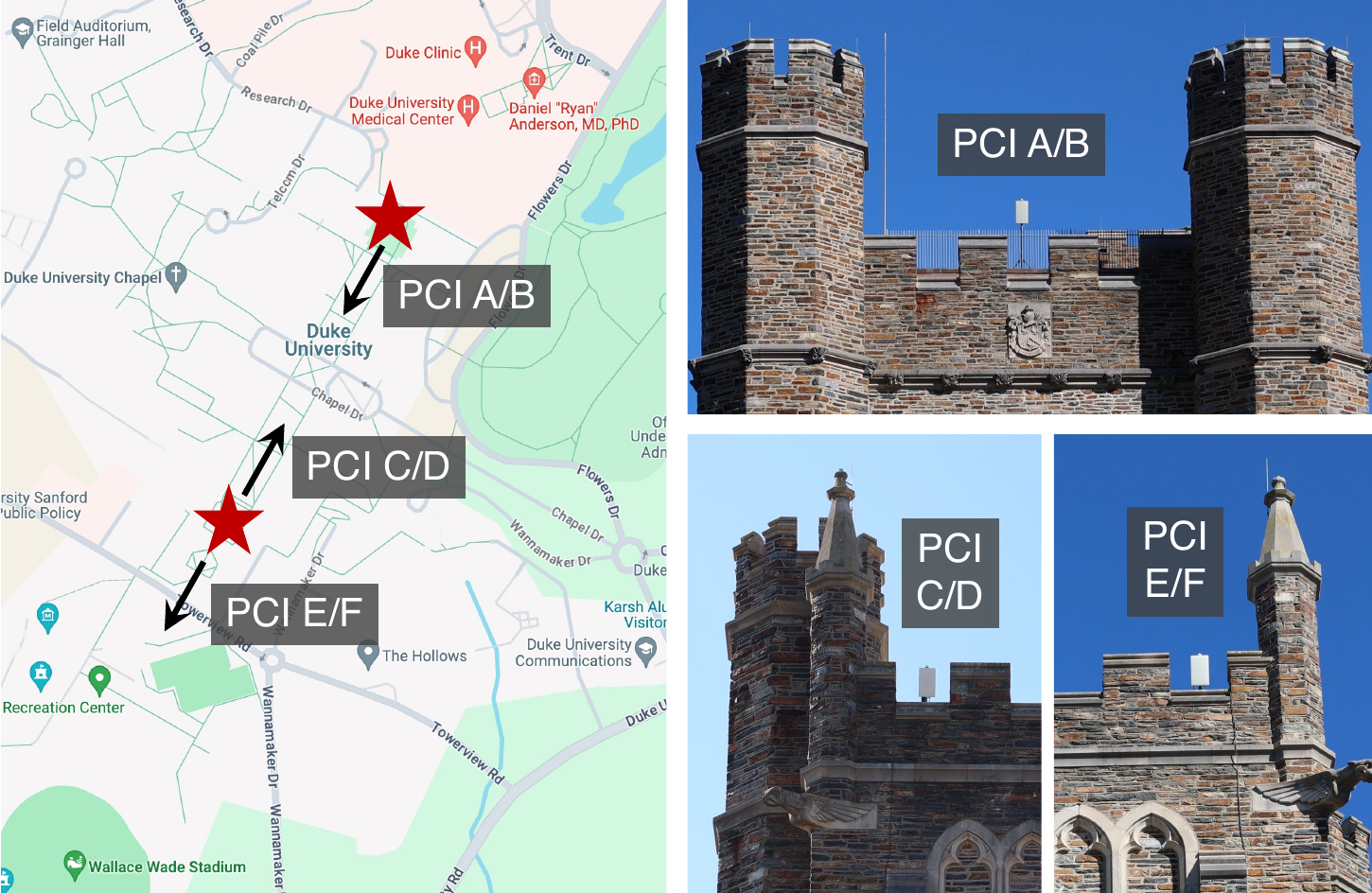}}
    \hspace{-2mm}
    \subfloat[]{
    \includegraphics[height=1.88in]{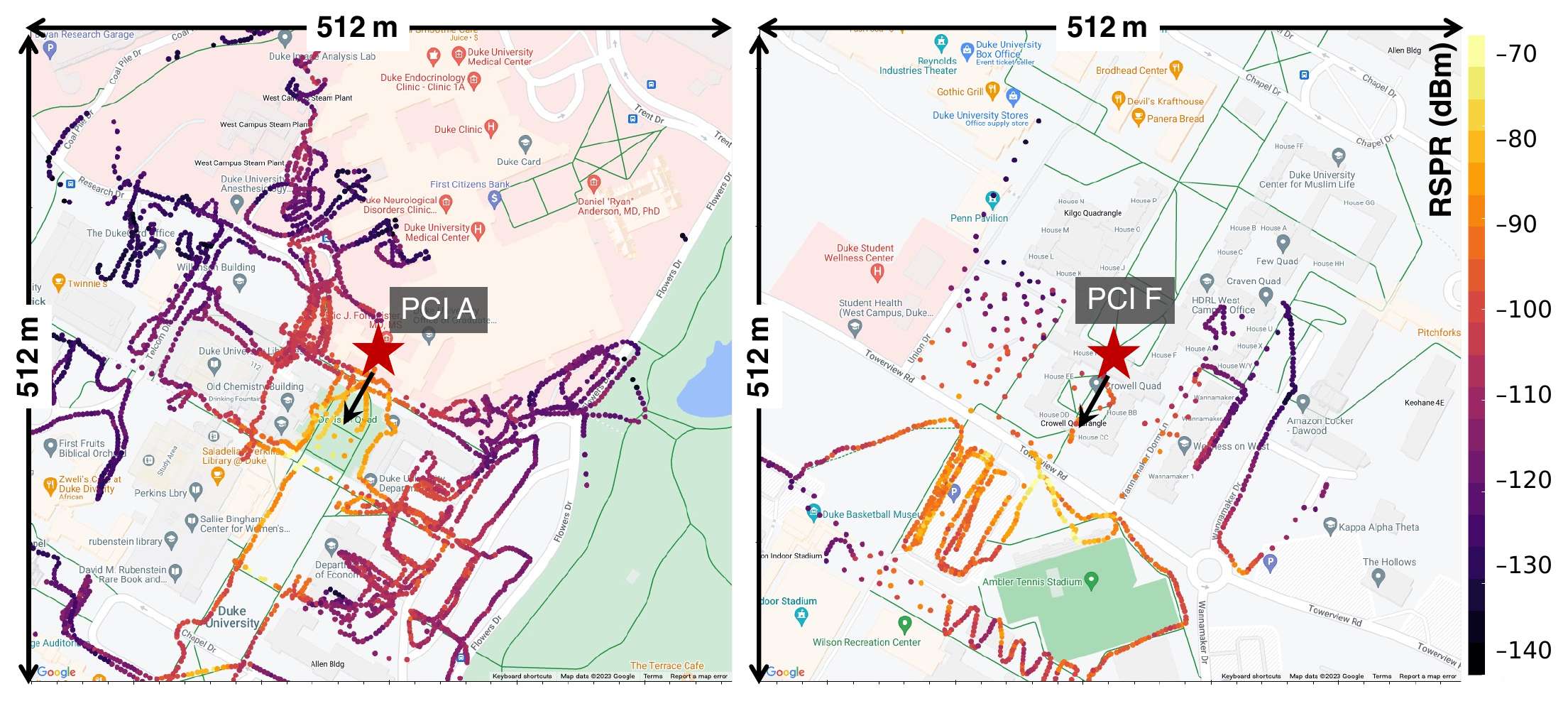}}
    \caption{(a) Six LTE cells operating in the CBRS band deployed on the Duke University West Campus (detailed cell information in Table~\ref{tab:cbrs_cell_info}).
    (b) Sample RSRP measurements collected by the UE when served by two cells (PCIs A and F) within the corresponding {512}\thinspace{m}$\times${512}\thinspace{m} area.}
    \label{fig:cbrs-cells-west-campus}
\end{figure*}

\subsection{Real-World Measurements}

\begin{table}[!t]
\centering
\caption{Specifications of six CBRS LTE cells (PCI A--F).}
\label{tab:cbrs_cell_info}
\vspace{-1mm}
\scriptsize
\begin{tabular}{|c|c|c|c|c|c|c|}
\hline
\textbf{\specialcell{PCI}} &\textbf{\specialcell{Carrier \\ Frequency}} & \textbf{Bandwidth} & \textbf{Height}    & \textbf{\specialcell{Azimuth \\ Orientation}}  & \textbf{\specialcell{Downtilt}} \\ 
\hline
\textbf{A} & 3.69\thinspace{GHz} & 20\thinspace{MHz} & 84.6\thinspace{ft} & 216$^{\circ}$ & 10$^{\circ}$ \\
\textbf{B} & 3.64\thinspace{GHz} & 20\thinspace{MHz} & 84.6\thinspace{ft} & 216$^{\circ}$ & 10$^{\circ}$ \\
\textbf{C} & 3.58\thinspace{GHz} & 20\thinspace{MHz} & 99.6\thinspace{ft} & 30$^{\circ}$ & 18$^{\circ}$ \\
\textbf{D} & 3.56\thinspace{GHz} & 20\thinspace{MHz} & 99.6\thinspace{ft} & 30$^{\circ}$ & 18$^{\circ}$ \\
\textbf{E} & 3.69\thinspace{GHz} & 20\thinspace{MHz} & 99.6\thinspace{ft} & 212$^{\circ}$ & 16$^{\circ}$ \\
\textbf{F} & 3.64\thinspace{GHz} & 20\thinspace{MHz} & 99.6\thinspace{ft} & 212$^{\circ}$ & 16$^{\circ}$ \\
\hline
\end{tabular}
\end{table}

\begin{table}[!t]
\centering
\caption{Total number of measurements collected by each UE type from different cells used for performance evaluation.}
\label{tab:measurement_number}
\vspace{-1mm}
\scriptsize
\begin{tabular}{|c|c|c|c|c|c|c|}
\hline
\textbf{PCI} & \textbf{A} & \textbf{B} & \textbf{C} & \textbf{D} & \textbf{E} & \textbf{F} \\ 
\hline
\textbf{Galaxy A42} & 4,297 & 3,461 & 1,599 & 1,436 & 529 & 970 \\ 
\textbf{Pixel 5} & 1,352 & 1,148 & 971 & 898 & 922 & 1,121 \\ 
\textbf{RPi w/ LTE HAT} & 4,110 & 8,469 & 3,741 & 4,425 & 1,792 & 3,915 \\ 
\hline
\textbf{Total \# of Meas.} & 9,759 & 13,078 & 6,311 & 6,759 & 3,243 & 6,006 \\ 
\hline
\end{tabular}
\end{table}

To evaluate the performance and generalizability of our proposed framework {\name}, we conducted a large-scale measurement campaign between 07/2023--11/2023, during which we collected user-side cellular data from six LTE cells operating in the CBRS band ({3.55--3.7}\thinspace{GHz}) deployed on the Duke University West Campus. Table~\ref{tab:cbrs_cell_info} summarizes the information about these LTE cells associated with (anonymized) PCI A through PCI F. As shown in Fig.~\ref{fig:cbrs-cells-west-campus}(a), two cells (PCI A/B) are deployed on top of the Davison Building at a height of {84.6}\thinspace{ft}, and four cells (PCI C/D/E/F) are deployed on top of the Crowell Quad House at a height of {99.6}\thinspace{ft}.
We select two {512}\thinspace{m}$\times${512}\thinspace{m} areas with PCI A/B and PCI C/D/E/F located at the center, respectively, as shown in Fig.~\ref{fig:cbrs-cells-west-campus}(b). One building map and 3D mesh are generated for each area, which are used by Sionna to obtain the ray tracing-based SS maps. \emph{Note that these two areas are not included in the training set (see Fig.~\ref{fig:area-north-america})},

We use three types of device serving as the UE, including three Samsung Galaxy A42 phones, one Google Pixel 5 phone, and one Raspberry Pi (RPi) 4B with an LTE HAT based on the Quectel EM060K-GL module. For the mobile phones, we develop a customized Android app to record the BS-side and user-side information including GPS coordinates, PCI, and reference signal receive power (RSRP), with a time resolution of 2 seconds. For the RPi setup, we develop customized Python-based scripts using AT commands~\cite{ATCommandsSet} to record similar information, with a time resolution of 0.5 seconds. Table~\ref{tab:measurement_number} summarizes the total number of collected measurements for each UE type when served by different cells, and Fig.~\ref{fig:cbrs-cells-west-campus}(b) shows two example of the collected RSRP measurements when the UE is served by two cells (PCI A/F).

We focus on the RSRP information collected by the UE when served by different cells (PCIs), which indicates the received SS observed by the UE at different locations. We average all the measurements within an area of $\resolution(\textrm{m}) \times \resolution(\textrm{m})$ to obtain the RSRP value corresponding to the pixel in the SS map, and the pixels without any measurements are excluded when calculating the prediction errors. We denote by $\mapRSRP$ the \emph{RSRP map} obtained from real-world measurements, and by $\mapRSRPSparse$ as the \emph{sparse RSRP map}, which contain a subset of RSRP measurement points in $\mapRSRP$. Using $\mapRSRPSparse$, $\mapBuilding$, and $\mapPathGainUMa$ as the inputs to the cascaded U-Net model, which is pre-trained using only the synthetic datasets, {\name} will predict the RSRP map for the entire area, $\mapRSRPEst$.

\subsection{Baseline Methods}
\label{ssec:implementation-baselines}

We evaluate the performance of {\name} against five baseline methods, including three analytical/statistical channel models, a ray tracing model utilizing Sionna, and PLNet~\cite{HuaweiKDD}, which is an ML-based radio propagation model based on CNNs. We convert the path loss values generated by these models into PG values by applying a multiplication factor of $-${1}. Below, we provide details about each baseline method.

\vspace{1.0ex}
\noindent\textbf{Friis free space model.}
The Friis transmission equation~\cite{Friis} is a fundamental formula that estimates the path gain, $\pathGain_{\textrm{Friis}}$\thinspace(dB), as a function of the link distance, $d$\thinspace(km), and the carrier frequency, $f$\thinspace(MHz), given by
\begin{align}
    \pathGain_{\textrm{Friis}}\thinspace[\textrm{dB}] = - \big[ 32.45 + 20\log_{10}(d) + 20\log_{10}(f) \big].
\end{align}

\vspace{1.0ex}
\noindent\textbf{3GPP urban macro (UMa) model.}
We also consider the urban marco (UMa) model specified by 3GPP TR38.901~\cite{3GPPTR38.901}, which includes both the line-of-sight (LOS) and the non-line-of-sight (NLOS) scenarios. For the LOS case, the path gain, $\pathGain_{\textrm{UMa-LOS}}$, as a function of the link distance, $d$\thinspace(m), and carrier frequency, $f$\thinspace(GHz), is given by
\begin{align}
    & \hspace{-2mm}
    \pathGain_{\textrm{UMa-LOS}}\thinspace[\textrm{dB}] = \nonumber \\
    & \hspace{0mm}
    \begin{cases}
    -\big[ 28.0 + 22 \log_{10}(d) + 20 \log_{10}(f) \big], & \text{if}~ d < d_{\textrm{BP}},
    \vspace{1mm} \\
    -\big[ 28.0 + 40 \log_{10}(d) + 20 \log_{10}(f) \\
    \hspace{5mm} - 9 \log_{10} ((d_{\textrm{BP}})^{2} + (h_{\textrm{TX}} - h_{\textrm{RX}})^2) \big], & \text{if}~ d \geq d_{\textrm{BP}},
    \end{cases}
\end{align}
where $d_{\textrm{BP}} = 4 (h_{\textrm{TX}}-1)(h_{\textrm{RX}}-1) \cdot f/c$ denotes the breakpoint distance and $c$ is the speed of light.
For the NLOS case, the path gain $\pathGain_{\textrm{UMa-NLOS}}$ is given by
\begin{align}
    \pathGain_{\textrm{UMa-NLOS}}\thinspace[\textrm{dB}] = \min \{ \pathGain_{\textrm{UMa-LOS}},\ \pathGain_{\textrm{UMa-NLOS}}' \},
\end{align}
where the second term is given by $\pathGain_{\textrm{UMa-NLOS}}'\thinspace[\textrm{dB}] = - \big[ 13.45 + 39.08\log_{10}(d) + 20 \log_{10}(f) - 0.6 (h_{\textrm{RX}} - 1.5) \big]$.

\vspace{1.0ex}
\noindent\textbf{Ericsson channel model.}
Another well-known and widely used channel model is the Ericsson model~\cite{Ericsson}, which is developed by Ericsson based on the modified Okumura-Hata model and supports varying parameters according to the propagation environment. The path gain under this model, $\pathGain_{\textrm{Eric}}$\thinspace(dB), as a function of the link distance, $d$ (km), carrier frequency, $f$\thinspace(MHz), TX antenna height, $h_{\textrm{TX}}$\thinspace(m), and RX antenna height, $h_{\textrm{RX}}$\thinspace(m), is given by
\begin{align}
    \pathGain_{\textrm{Eric}}\thinspace[\textrm{dB}]
    & = - \big[ a_{0} + a_{1}\log_{10}(d) + a_{2}\log_{10}(h_{\textrm{TX}}) \nonumber \\
    & \hspace{8mm} + a_{3}\log_{10}(h_{\textrm{TX}}) \cdot \log_{10}(d) \nonumber \\
    & \hspace{8mm} - 3.2 (\log_{10}(11.75 h_{\textrm{RX}})^{2}) + g(f) \big],
\end{align}
where $g(f) = 44.49 \log_{10}(f) - 4.78 (\log_{10}(f))^{2}$.
Since we focus on urban environments, the coefficients $a_{0}$, $a_{1}$, $a_{2}$, and $a_{3}$ are set as 36.2, 30.2, 12, and 0.1, respectively~\cite{Ericsson}.

\vspace{1.0ex}
\noindent\textbf{Ray tracing model utilizing Sionna.}
We implement another baseline method based on pure ray tracing employing Sionna~0.15.1. In particular, we follow the procedure described in Sections~\ref{ssec:system-design-building-maps} and~\ref{ssec:system-design-ray-tracing} with the ray tracing parameters in Table \ref{tab:SionnaSpec} and the cell information in Table~\ref{tab:cbrs_cell_info} to generate the PG map. Then, the SS map can be obtained by plugging the antenna parameters into the link budget equation {\eqref{eq:signal-strength-link-budget}.

\vspace{1.0ex}
\noindent\textbf{PLNet.}
PLNet~\cite{HuaweiKDD} is a state-of-the-art ML-based method that leverages detailed environmental data and cell specifications to accurately predict SS maps. In particular, PLNet uses a single U-Net architecture, whose input features include building, terrain, and clutter maps as well as antenna height, orientation, and beam pattern information. The U-Net model is trained using synthetic SS maps generated by the Siradel SAS software~\cite{Siradel}, which is a commercial \emph{licensed} ray tracing tool for accurate signal propagation modeling. Due to the proprietary nature of Siradel SAS, our implementation of PLNet utilizes the synthetic directional PG datasets generated by Sionna and the same U-Net model architecture as described in~\cite{HuaweiKDD}, where the terrain and clutter maps are excluded from the input features.

\vspace{1.0ex}
\noindent\textbf{Model calibration.}
We convert the PG values generated by each baseline method into their respective SS maps based on {\eqref{eq:signal-strength-link-budget}}. To compensate for uncertain system parameters, such as the UE orientation and antenna gain, we introduce a constant offset when calculating the SS maps for each baseline. These offsets are optimized using 100 data points collected from our measurement campaign, which enable calibration and fine-tuning of each baseline and ensure fair comparison across different methods. The performance of each calibrated method is then evaluated using real-world measurement data.

\section{Evaluation}
\label{sec:evaluation}

We evaluate the performance of {\name} using real-world measurements and compare it with various baseline methods. We focus on the RSRP values recorded by the UE serving as the SS maps, and our proposed framework can also be extended to mapping other key metrics such as SINR and RSSI, as described in Section~\ref{sec:system-design}.

\begin{figure}[!t]
    \centering
    \includegraphics[width=0.48\columnwidth]{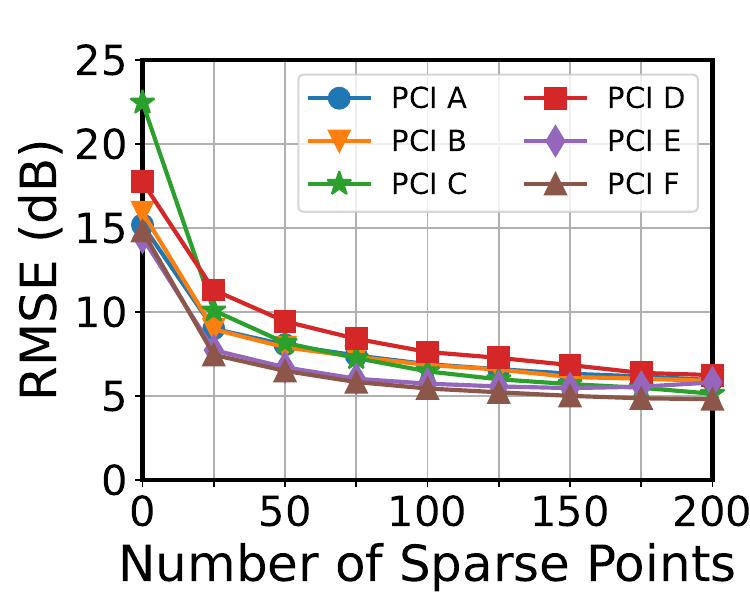}
    \includegraphics[width=0.48\columnwidth]{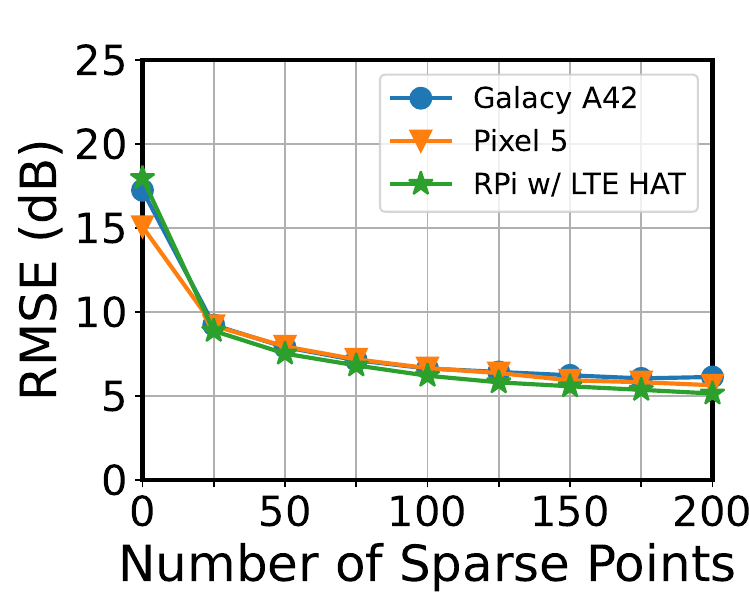}
    \vspace{-1mm}
    \caption{Root-mean-square error (RMSE) of the RSRP values predicted by {\name} with varying number of measurement points in the sparse map for different PCIs (\emph{left}) and device types (\emph{right}).}
    \label{fig:impact-sparse-map-size}
\end{figure}

\vspace{1.0ex}
\noindent\textbf{Selecting the sparse SS map size.}
We first explore the answer to following question:
How many RSRP measurement points from a given area are sufficient for the cascaded U-Net model, pre-trained using only synthetic datasets, to accurately predict the full RSRP map for the entire area?
Fig.~\ref{fig:impact-sparse-map-size} shows the root-mean-square error (RMSE) of the predicted SS map for the entire area, $\mapRSRPEst$, compared to the ground truth SS map, $\mapRSRP$, with varying sparse SS map sizes, i.e., sparse SS maps comprising different numbers of measurement points are used to predict the full SS map.

The results show that with only 50 measurement points, {\name} achieves an average RMSE of {7.79}\thinspace{dB} across all PCIs and UE types. The RMSE is further improved to {6.50}\thinspace{dB} and {5.64}\thinspace{dB} with 100 and 200 measurement points, respectively. Based on these results, we empirically select to use sparse SS maps with 100 measurement points for the prediction of the full SS maps to evaluate the performance of {\name} and its comparison to other baseline methods.

\begin{figure}[!t]
    \centering
    \vspace{2.5mm}
    \includegraphics[width=0.96\columnwidth]{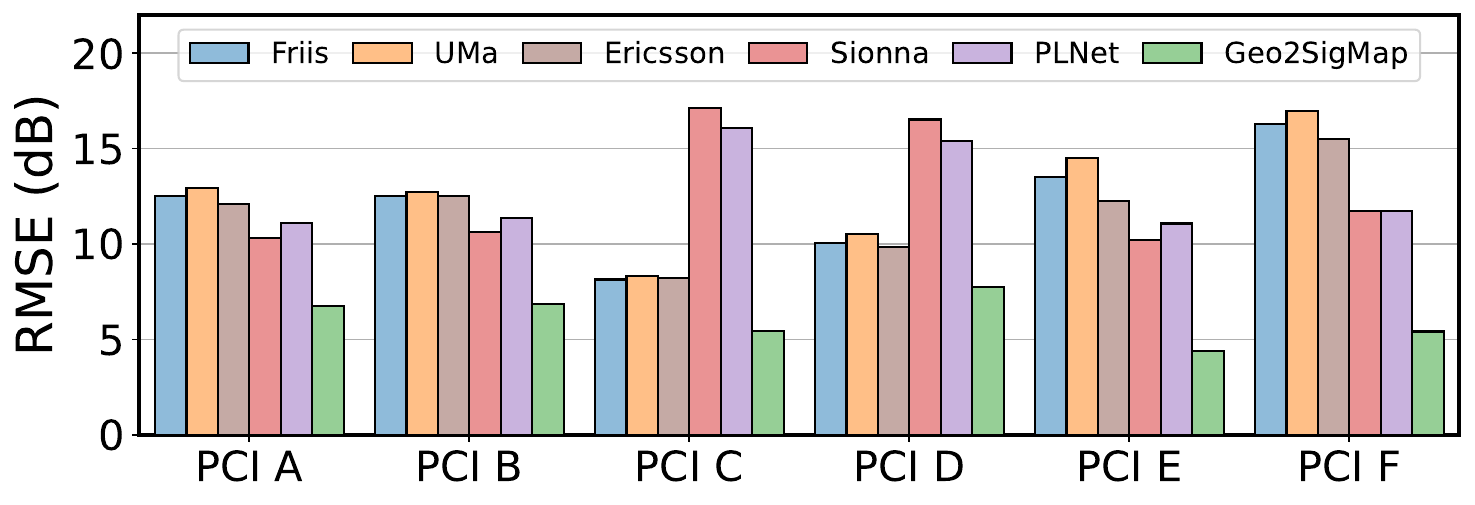}
    \vspace{-1mm}
    \caption{RMSE of the predicted RSRP maps for individual PCIs achieved by {\name} compared to the baselines.}
    \label{fig:bar-method}
\end{figure}

\vspace{1.0ex}
\noindent\textbf{RMSE of predicted RSRP maps.}
We then evaluate the performance of {\name} in terms of the predicted RSRP maps and compare it with five baselines, including three statistical channel models (Friis, UMa, and Ericsson), Sionna-based ray tracing, and PLNet, as described in Section~\ref{ssec:implementation-baselines}. Note that the analytical/statistical channel models and Sionna-based ray tracing models are calibrated with an offset optimized using 100 measurement points. In contrast, PLNet and {\name} are trained using only synthetic datasets.
Fig.~\ref{fig:bar-method} shows the RMSE of the RSRP maps predicted for individual PCIs by {\name} and various baseline methods, where the measurements collected across all UE types are aggregated for each PCI. In general, the RMSE values achieved by different methods exhibit similarity for each pair of two PCIs sharing the same configurations (i.e., PCI A/B, C/D, and E/F, see Table~\ref{tab:cbrs_cell_info}). The results show that {\name} consistently outperforms all baseline methods without relying on any measurements: it achieves an RMSE between {4.39--7.74}\thinspace{dB} across the six PCIs, representing an average RMSE improvement of {3.59}\thinspace{dB} compared to the next best performing method.

\begin{figure}[!t]
    \centering
    \vspace{1.8mm}
    \subfloat[Samsung Galaxy A42]{
    \includegraphics[width=0.98\columnwidth]{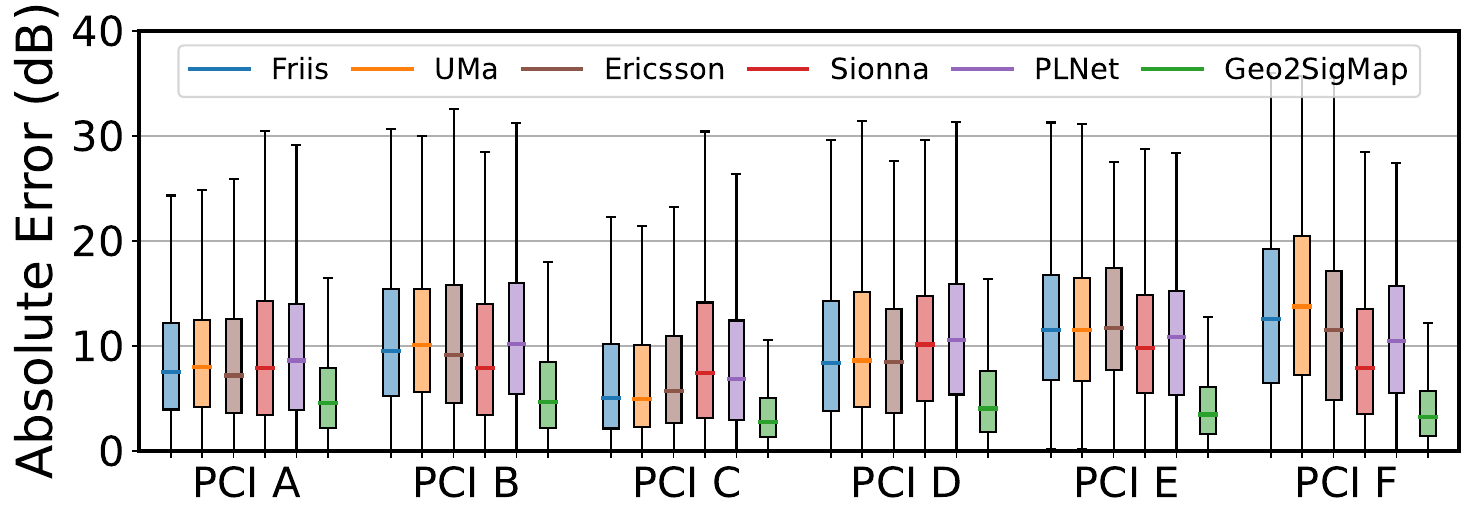}}
    \vspace{-3mm} \\
    \subfloat[Google Pixel 5]{
    \includegraphics[width=0.98\columnwidth]{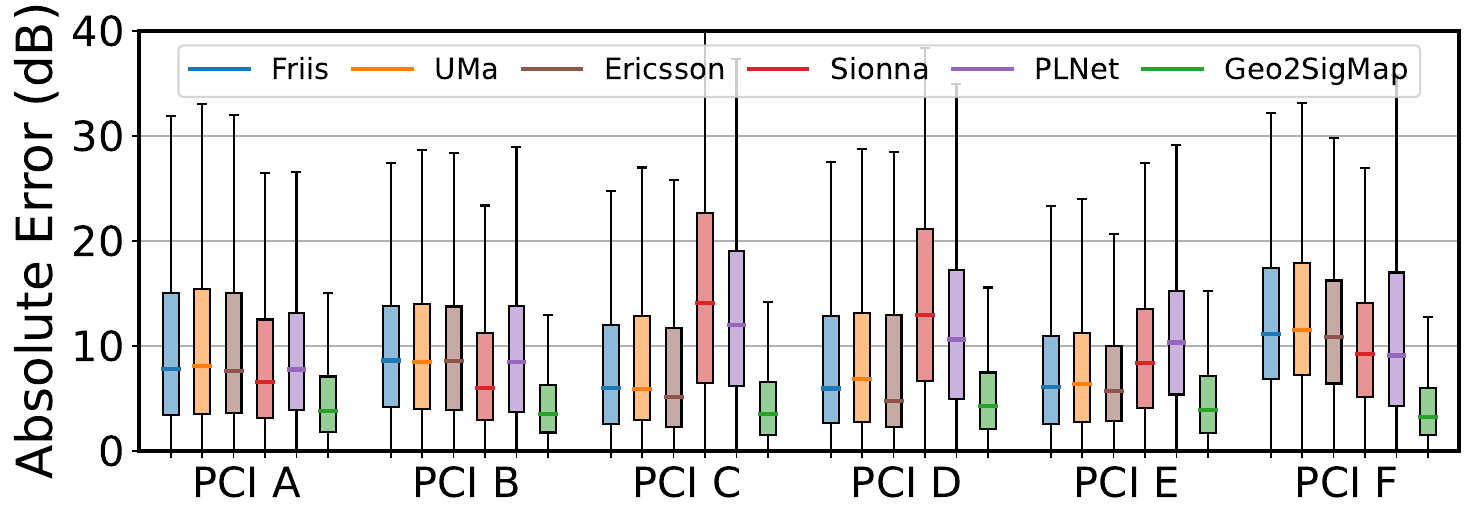}}
    \vspace{-3mm} \\
    \subfloat[Raspberry Pi with LTE HAT.]{
    \includegraphics[width=0.98\columnwidth]{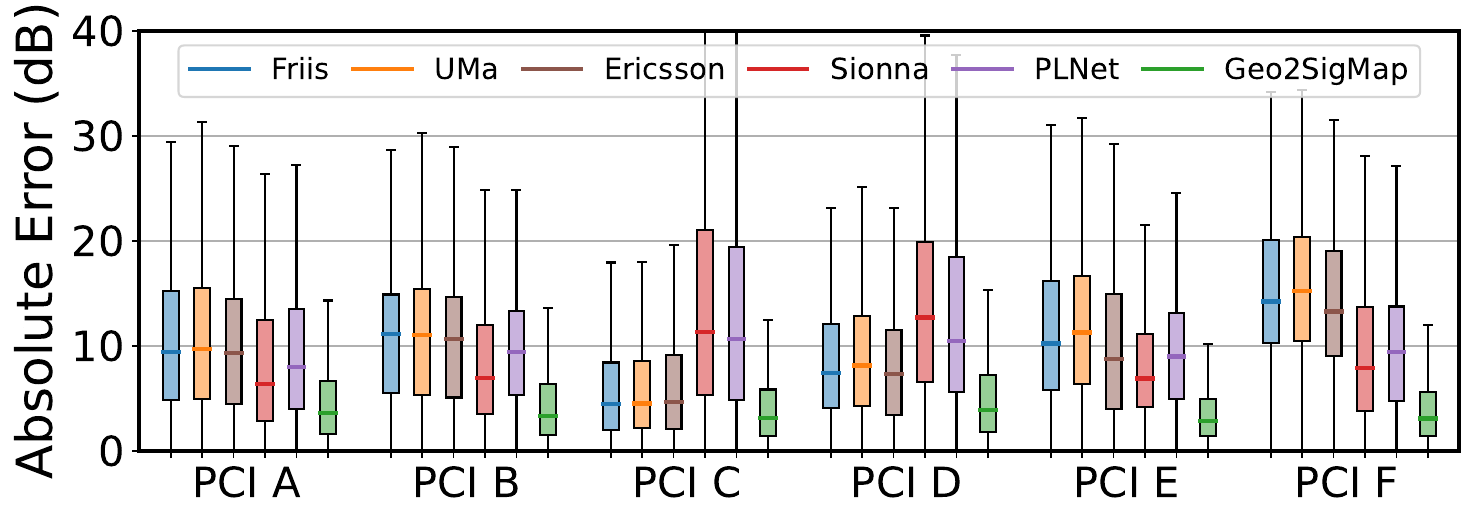}}
    \caption{Absolute error of the predicted RSRP maps for individual PCIs achieved by {\name} compared to the baseline methods and categorize by three UE types.}
    \label{fig:box-method}
    \vspace{-3mm}
\end{figure}

\vspace{1.0ex}
\noindent\textbf{Error distribution of predicted RSRP maps.}
Fig.~\ref{fig:box-method} presents the absolute error of the RSRP maps predicted for individual PCIs by {\name} and the baseline methods, 
displayed in the standard boxplot format and categorized by three UE types. The results show that {\name} achieves significantly improved prediction accuracy in both median and tail performance metrics. Specifically, {\name} achieves an average median absolute error of the predicted RSRP maps between {3.77/3.69/3.31}\thinspace{dB} across the six PCIs for the Galaxy~A42, Pixel~5, and RPi UE type, respectively. This represents an average improvement of {4.62/3.34/5.14}\thinspace{dB} compared to the second best baseline method.
In addition, compared to the baseline methods, {\name} achieves a much smaller average interquartile range (IQR) of the absolute error of {5.06/5.04/4.58}\thinspace{dB} across the six PCIs for the three UE types, respectively, which is {5.38/4.78/4.49}\thinspace{dB} lower than that of the cloest competing baseline method.
Furthermore, Fig.~\ref{fig:stack-hist-narrow} shows the 
probability distribution functions (PDFs) of the RSRP prediction errors achieved by {\name}, aggregated across six PCIs and categorized by the three UE types. The results reveal that our proposed method achieves a more focused error distribution with smaller variations, outperforming all the baseline methods.

\begin{figure}[!t]
    \vspace{-3mm}
    \centering
    \subfloat[Samsung Galaxy A42]{
    \includegraphics[width=0.33\columnwidth]{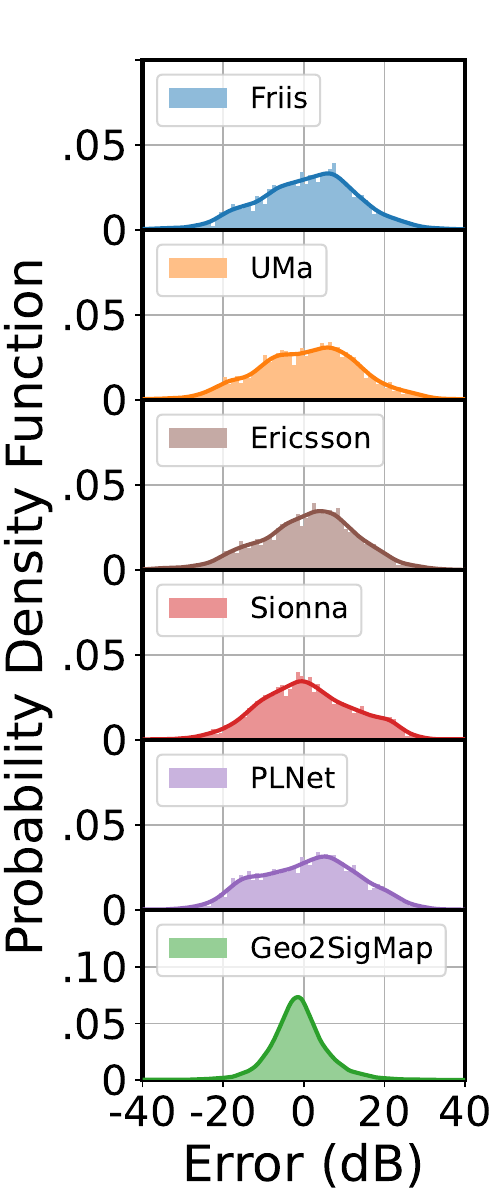}}
    \hspace{-2mm}
    \subfloat[Google Pixel 5]{\includegraphics[width=0.33\columnwidth]{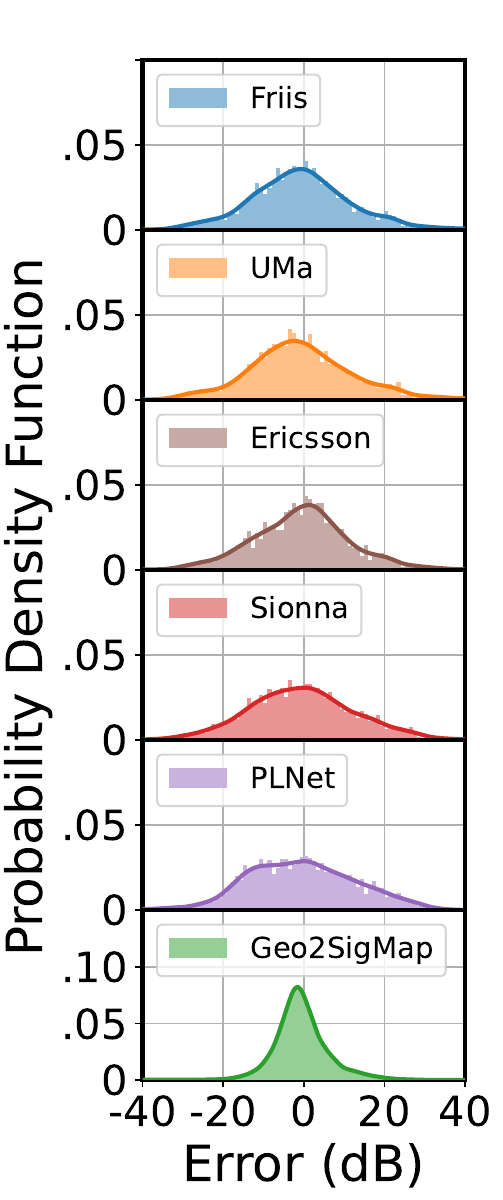}}
    \hspace{-2mm}
    \subfloat[RPi w/ LTE HAT]{
    \includegraphics[width=0.33\columnwidth]{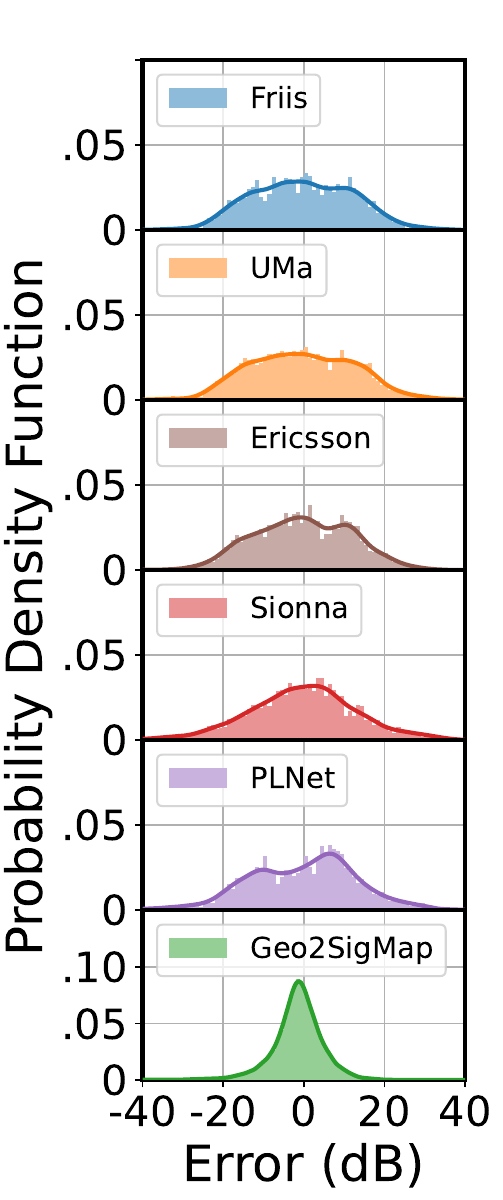}}
    \caption{Error distribution of the predicted RSRP maps achieved by {\name} compared to the baseline methods.}
    \label{fig:stack-hist-narrow}
\end{figure}

\section{Discussions and Limitations}
\label{sec:discussions}

\noindent\textbf{Impact of other environmental factors.}
In addition to 3D buildings, PG can be affected by other environmental factors, such as the terrain profile, foliage condition, and human/vehicle blockage. Therefore, there is room for improvement by incorporating a more diverse set of objects into the ray tracing tool, which can provide more accurate results for ray tracing. 
Fortunately, Google has recently released the ``3D Tile API''~\cite{Google_3D} on Google I/O 2023, which includes not only building maps but also foliage condition and terrain profiles. Currently, this API covers over 2,500 cities across 49 countries worldwide. 
We plan to investigate the integration of this API into our developed pipeline to further improve the environmental awareness during the ray tracing process.

\vspace{1.0ex}
\noindent\textbf{Model adaptation on different geographic and network settings.}
So far, our synthetic dataset generation, model training, and performance evaluation focus on the CBRS band ({3.55--3.7}\thinspace{GHz}) with an area dimension of {512}\thinspace{m}$\times${512}\thinspace{m}.
As a result, our model may exhibit degraded performance with different area sizes (e.g., {1,024}\thinspace{m}$\times${1,024}\thinspace{m})
and at different carrier frequency bands used by commercial cellular networks.
To address these challenges, we can generate a dedicated synthesized dataset that accounts for these factors and investigate transfer learning techniques to generalize pre-trained cascaded U-Net model to different scenarios.
\section{Conclusion}
\label{sec:conclusion}

We presented the design of {\name}, an efficient framework for high-fidelity RF signal mapping leveraging geographic databases and a novel cascaded U-Net model.
We first developed an automated pipeline that efficiently generates 3D building and path gain maps via the integration of a suite of open-sourced tools, including OSM, Blender and Sionna. Then, the cascaded U-Net model pre-trained on synthetic datasets utilizes the building map and sparse SS map as input to predict the full SS map for the target (unseen) area.
We extensively evaluated the performance of {\name} using large-scale field measurement collected using three UE types across six CBRS LTE cells deployed on the Duke University West Campus. Our results showed that {\name} achieves significantly improved RMSE of the SS map prediction compared to existing baseline methods.

\section*{Acknowledgments}
The work was supported in part by NSF grants CNS-2112562, CNS-2128638, CNS-2211944, and AST-2232458. We thank Joe Scarangella (RF Connect), and John Board, William Brockelsby, and Robert Johnson (Duke University) for their contributions to this work. 

\bibliographystyle{IEEEtran}
\bibliography{reference}

\end{document}